\documentclass{PoS}

\usepackage[utf8]{inputenc}
\usepackage{amsmath}
\usepackage{amssymb}
\usepackage{xspace}
\usepackage{tikz}

\newcommand{\ket}[1]{\ensuremath{|#1\rangle}\xspace}
\newcommand{\bra}[1]{\ensuremath{\langle #1|}\xspace}

\newcommand{\g}[0]{\ensuremath{\gamma}}

\title{Basic properties of GPDs and modelling of the latter}

\ShortTitle{Basic properties of GPDs}

\author{\speaker{Cédric Mezrag}
  \\
        IRFU, CEA, Université Paris-Saclay, F-91191 Gif-sur-Yvette, France\\
        E-mail: \email{cedric.mezrag@cea.fr}}

\author{Nabil Chouika\\
       IRFU, CEA, Université Paris-Saclay, F-91191 Gif-sur-Yvette, France\\
       E-mail: \email{nabil.chouika@cea.fr}}
      
\author{Hervé Moutarde\\
       IRFU, CEA, Université Paris-Saclay, F-91191 Gif-sur-Yvette, France\\
       E-mail: \email{herve.moutarde@cea.fr}}

\author{Jose Rodr\`iguez-Quintero\\
       Department of Integrated Sciences and Center for Advanced Studies in Physics,\\
       Mathematics and Computation; University of Huelva, E-21071 Huelva; Spain.\\
       E-mail: \email{jose.rodriguez@dfaie.uhu.es}}

\abstract{We present here a new method based on the Radon transform to model Generalised Parton Distributions (GPDs). It allows to fulfil all theoretical constraints applying on GPDs, especially polynomiality and positivity at the same time. More specifically, we show how polynomiality can be systematically restored within the framework of the overlaps of Lightfront Wave Functions (LFWFs). It provides a systematic way to extend models defined solely in the DGLAP kinematical region to the ERBL one. We then exemplify our approach using LFWFs models.}

\FullConference{Light Cone 2019 - QCD on the light cone: from hadrons to heavy ions - LC2019\\
		16-20 September 2019\\
		Ecole Polytechnique, Palaiseau, France}

\begin{document}

\section{Introduction}

Since they have been introduced two decades ago \cite{Mueller:1998fv,Ji:1996nm,Radyushkin:1997ki}, Generalised Parton Distributions (GPDs) have been under deep theoretical and experimental studies. On the experimental side, we are entering the precision era \cite{Kumericki:2016ehc}, being sensitive to higher-twist and next-to-leading order effects \cite{Defurne:2017paw}. On the theory side, progresses have been also performed, allowing us to extract Form Factors of exclusive processes \cite{Moutarde:2018kwr,Moutarde:2019tqa,Kumericki:2019mgk} with a better control of uncertainties. Another breakthrough was made in the possibility to take into account higher-twist kinematical effects \cite{Braun:2012hq,Braun:2012bg} shown to be sizeable \cite{Defurne:2015kxq}. On top of this, new open source software like PARTONS \cite{Berthou:2015oaw} are now available.

Modelling GPDs remains however a challenge and various approaches have been developed. Phenomenologists mostly rely on Double Distributions \cite{Mueller:1998fv,Radyushkin:1997ki} and more precisely on the Radyushkin Ansatz \cite{Mukherjee:2002gb} (see for instance \cite{Goloskokov:2005sd,Mezrag:2013mya}) and on conformal space approaches \cite{Kumericki:2009uq,Kumericki:2015lhb}. Model based on dynamical assumptions of QCD are still not in sight for the nucleon, although some attempts had been performed in the meson sector \cite{Chang:2014lva,Mezrag:2014jka}. Lattice calculations of GPDs Mellin moments have also been performed \cite{Alexandrou:2013joa} and attempts to compute quasi-GPDs start to emerge \cite{Chen:2019lcm}.

Within this general picture, we present here briefly the new approach we developed in \cite{Chouika:2017dhe,Chouika:2017rzs} to allow the computation of GPD models fulfilling \emph{a priori} all the theoretical constraints required by the symmetries of QCD. Our framework allow notably to fulfil at the same time both positivity and polynomiality, which are known to be hard to conciliate together a priori.

\section{Basic Properties of GPDs}

GPDs are defined as the Fourier transform of a non-perturbative matrix element of a quark operators depending on a light-like distance. For a pseudo-scalar hadron (a pion), one has \cite{Diehl:2003ny}:
\begin{equation}
  \label{eq:GPDDef}
   H(x,\xi,t) = \frac{1}{2}\int_{-\infty}^{+\infty} \frac{\textrm{d}z^-}{2\pi} e^{ixP^+z^-}\left.
    \bra{P+\frac{\Delta}{2}}\bar{\psi}\left(-\frac{z}{2}\right)\g ^+ \psi\left(\frac{z}{2}\right)\ket{P-\frac{\Delta}{2}}
  \right|_{z^+=z_\perp = 0},
\end{equation}
where $P$ is the average momentum of the hadron, $\Delta$ the momentum transfer,  $x$ is the average momentum fraction carried by the active quark, the skewness $\xi = \Delta^+/(2P^+)$ represent half of the fraction of longitudinal momentum exchanged and $t = \Delta^2$ is the usual Mandelstam variable. Through this work, lightfront coordinate will be used : $v^\pm = (v^0 \pm v^3)/\sqrt{2}$. 

\subsection{Positivity and Lightfront Wave Functions}

The lightfront quantization allows one to expand hadron states on a Fock basis \cite{Brodsky:1997de}:
\begin{equation}
  \label{eq:FockBasis}
  \ket{H,p,\lambda} = \sum_{N,\beta} \int \textrm{d}x_1\dots \textrm{d}x_N \frac{\textrm{d}k^\perp_1\dots \textrm{d}k^\perp_N}{(16\pi^3)^{N-1}} \delta_x \delta_{k_\perp}\Psi_N(x_1,k^\perp_1\dots x_N,k^\perp_N) \ket{q_1\dots q_N},
\end{equation}
with $\delta_x = \delta(1-\sum_ix_i)$ and $\delta_{k_\perp} = \delta^{(2)}(P_\perp - \sum_ik^\perp_i)$ and $\Psi_N$ being the $N$-body lightfront wave-functions (LFWFs). This expansion provides one with a systematic way to compute GPDs in terms of overlaps of LFWFs \cite{Diehl:2000xz}. However, the type of overlap depends on the kinematical region investigated. Indeed for $|x|\ge|\xi|$ (the DGLAP region), the overlap involves LFWFs of the same number of partons (schematically $\sum_N \bar{\Psi}_N\Psi_N$) yielding a formal scalar product, and therefore the Cauchy-Schwartz inequality associated. This yields the positivity property \cite{Pire:1998nw}:
\begin{equation}
  \label{eq:Positivity}
  |H(x,\xi,t)| \le \sqrt{q\left(\frac{x-\xi}{1-\xi}\right)q(\left(\frac{x+\xi}{1+\xi} \right)},
\end{equation}
where $q$ is the pion PDF. However, in the $|x| \le |\xi|$ kinematical region, the initial state has two more partons than the final one, leading to an off-diagonal overlap formula (schematically $\sum_N \bar{\Psi}_{N-2}\Psi_N$). Therefore, a $N$-body truncation in the Fock basis \eqref{eq:FockBasis} yields an ambiguous description of the ERBL region, and breaks the polynomiality properties of GPDs described below.

\subsection{Polynomiality and Double Distributions}

The Mellin moments of GPDs need to be polynomials in $\xi$ due to Lorentz covariance:
\begin{equation}
  \label{eq:Polynomiality}
  M_n(\xi,t) = \int_{-1}^1 \textrm{d}x x^n H(x,\xi,t) = \sum_{i=0}^{\left[\frac{n}{2} \right]} c_{i,n}(t) \xi^{2i} + mod(n,2) \xi^{n+1}c_{n+1,n}(t).
\end{equation}
Introducing a function $D(x/\xi,t)\Theta(|\xi|-|x|)$ labelled the $D$-term such that:
\begin{equation}
  \label{eq:DtermDef}
  \int_{-|\xi]}^{|\xi|}\textrm{d}x~x^n D\left(\frac{x}{\xi},t \right) =  mod(n,2) \textrm{sign} (\xi)  \xi^{n+1}c_{n+1,n}(t),
\end{equation}
one can directly find the Lugwig-Helgason consistency condition:
\begin{equation}
  \label{eq:LudwigHelgason}
   \int_{-1}^1 \textrm{d}x x^n \left(H(x,\xi,t)- \textrm{sign}(\xi)\Theta\left(|\xi|-|x| \right) D\left(\frac{x}{\xi},t \right)\right) = \sum_{i=0}^{\left[\frac{n}{2} \right]} c_{i,n}(t) \xi^{2i},
\end{equation}
and consequently, $H-D$ is the Radon transform of another object, called the Double Distribution\footnote{The presentation is done in the Polyakov-Weiss DD scheme, but is valid for all DD scheme \cite{Chouika:2017dhe}.} (DD). The latter were originally introduced in \cite{Mueller:1998fv,Radyushkin:1997ki} as the 2D Fourier transform of non-local matrix element. It is clear here that the polynomiality property and the existence of DDs are equivalent, and contrary to what may have been thought previously, DD are not simply a convenient way to fulfil the polynomiality property. The Radon transform relation was first noticed in \cite{Teryaev:2001qm} and exploit for instance beyond GPD physics in \cite{Anikin:2019oes}.

\section{Inverse Radon Transform}

The Radon transform admit an inverse transform, which can be used to compute DD from GPDs. However, proceeding naively, one would need to know the GPDs not only for $\xi\in[-1,1]$ but for $\xi \in \mathbb{R}$, i.e. including the Generalised Distribution Amplitude kinematical sector, \emph{a priori} precluding useful applications of the inverse Radon transform. We showed in \cite{Chouika:2017dhe} that the situation is actually much better. Using the Boman and Todd-Quinto theorem \cite{boman:1987rad}, one can prove that the DGLAP region is sufficient to entirely constraint the Double Distributions up to $D$-term like singularities. This generalises previous results obtain within specific functional forms \cite{Hwang:2007tb,Mueller:2014tqa}

\begin{figure}[t]
  \centering
        \begin{tikzpicture}
        \tikzstyle{ArrowRed} = [ ->, very thick, red, >=latex ] ;
        \tikzstyle{ArrowGreen} = [ ->, very thick, teal, >=latex ] ;
        \tikzstyle{ArrowBlack} = [ ->, very thick, black, >=latex ] ;
        %
        \draw[ rounded corners ] ( 0., 2.25 ) rectangle ( 2.5, 4. ) ;
        \node at ( 1.25,3.125 ) {\begin{minipage}{3.25cm}{\begin{center}\textcolor{blue}{Lightfront \\ 
                  Wave \\Functions}\end{center}}\end{minipage}} ;
          \draw[ rounded corners ] ( 3., 2.25 ) rectangle ( 5.5, 4. ) ;
          \node at ( 4.25,3.125 ) {\begin{minipage}{3.25cm}{\begin{center}\textcolor{blue}{GPDs in the \\ 
                    DGLAP region}\end{center}}\end{minipage}} ;
          \draw[ rounded corners ] ( 6., 2.25 ) rectangle ( 8.5, 4. ) ;
          \node at ( 7.25,3.125 ) {\begin{minipage}{3.25cm}{\begin{center}\textcolor{blue}{Double \\ 
                    Distributions}\end{center}}\end{minipage}} ;
          \draw[ rounded corners ] ( 9., 2.25 ) rectangle ( 11.5, 4. ) ;
          \node at ( 10.25,3.125 ) {\begin{minipage}{3.25cm}{\begin{center}\textcolor{blue}{GPDs in the \\ 
                    ERBL region}\end{center}}\end{minipage}} ;
          \draw[ rounded corners ] ( 1.625, 0. ) rectangle ( 3.875, 1.75) ;
          \node at ( 2.75, 0.875 ) {\begin{minipage}{3.25cm}{\begin{center}\textcolor{black}{Overlap \\ 
                    of LFWFs}\end{center}}\end{minipage}} ;
          \draw[ rounded corners ] ( 4.625, 0. ) rectangle ( 6.875, 1.75 ) ;
          \node at ( 5.75, 0.875 ) {\begin{minipage}{3.25cm}{\begin{center}\textcolor{black}{Inverse \\ 
                    Radon\\ Transform}\end{center}}\end{minipage}} ;
          \draw[ rounded corners ] ( 7.625, 0. ) rectangle ( 9.875, 1.75 ) ;
          \node at ( 8.75, 0.875 ) {\begin{minipage}{3.25cm}{\begin{center}\textcolor{black}{Radon\\ Transform}\end{center}}\end{minipage}} ;
        %
          \draw[ rounded corners ] ( 1.625, 4.25 ) rectangle ( 3.875, 5.25) ;
          \node at ( 2.75, 4.75 ) {\begin{minipage}{3.25cm}{\begin{center}\textcolor{black}{Positivity\\
fulfilled}\end{center}}\end{minipage}} ;
          \draw[ rounded corners ] ( 7.625, 4.25 ) rectangle ( 9.875, 5.25 ) ;
          \node at ( 8.75, 4.75 ) {\begin{minipage}{3.25cm}{\begin{center}\textcolor{black}{Polynomiality\\ fulfilled}\end{center}}\end{minipage}} ;
        %

          \draw[ ArrowGreen ] ( 2.75, 1.75 ) to ( 2.75, 3. ) ;
          \draw[ ArrowGreen ] ( 5.75, 1.75 ) to ( 5.75, 3. ) ;
          \draw[ ArrowGreen ] ( 8.75, 1.75 ) to ( 8.75, 3. ) ;
        

          \draw[ ArrowBlack ] ( 2.5, 3.125 ) to ( 3., 3.125 ) ;

          \draw[ ArrowBlack ] ( 5.5, 3.125 ) to ( 6., 3.125 ) ;

          \draw[ ArrowBlack ] ( 8.5, 3.125 ) to ( 9., 3.125 ) ;


          \draw[ ArrowRed ] ( 2.75, 3.25 ) to ( 2.75, 4.25 ) ;

          \draw[ ArrowRed ] ( 8.75, 3.25 ) to ( 8.75, 4.25 ) ;

      \end{tikzpicture}
  \caption{Modelling strategy for GPDs fulfilling all theoretical constraints.}
  \label{fig:Strategy}
\end{figure}
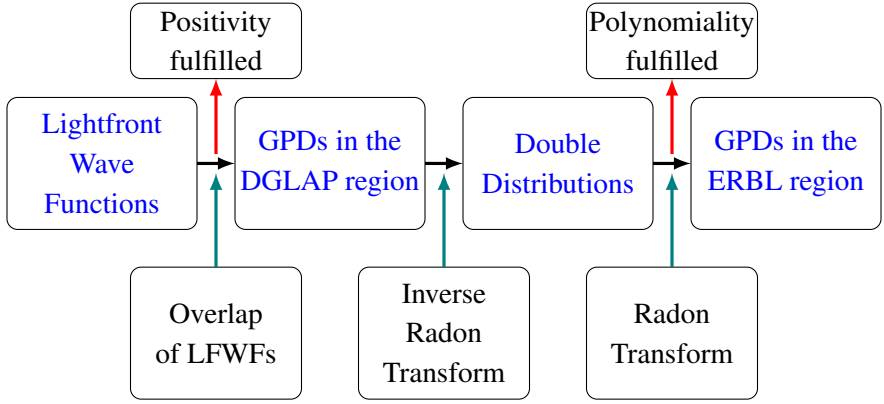

This result is crucial, as it allows to develop a new modelling strategy, combining the best of the LFWFs and DDs worlds to fulfil \emph{a priori} both the positivity and polynomiality properties. The strategy consists in computing the DGLAP region from LFWFs, and then computing the DDs from the DGLAP region only. This way LFWFs will ensures positivity while DDs polynomiality. A sketch of the strategy is shown on figure \ref{fig:Strategy}.

\begin{figure}[b]
  \centering
  \includegraphics[width = 0.4\textwidth]{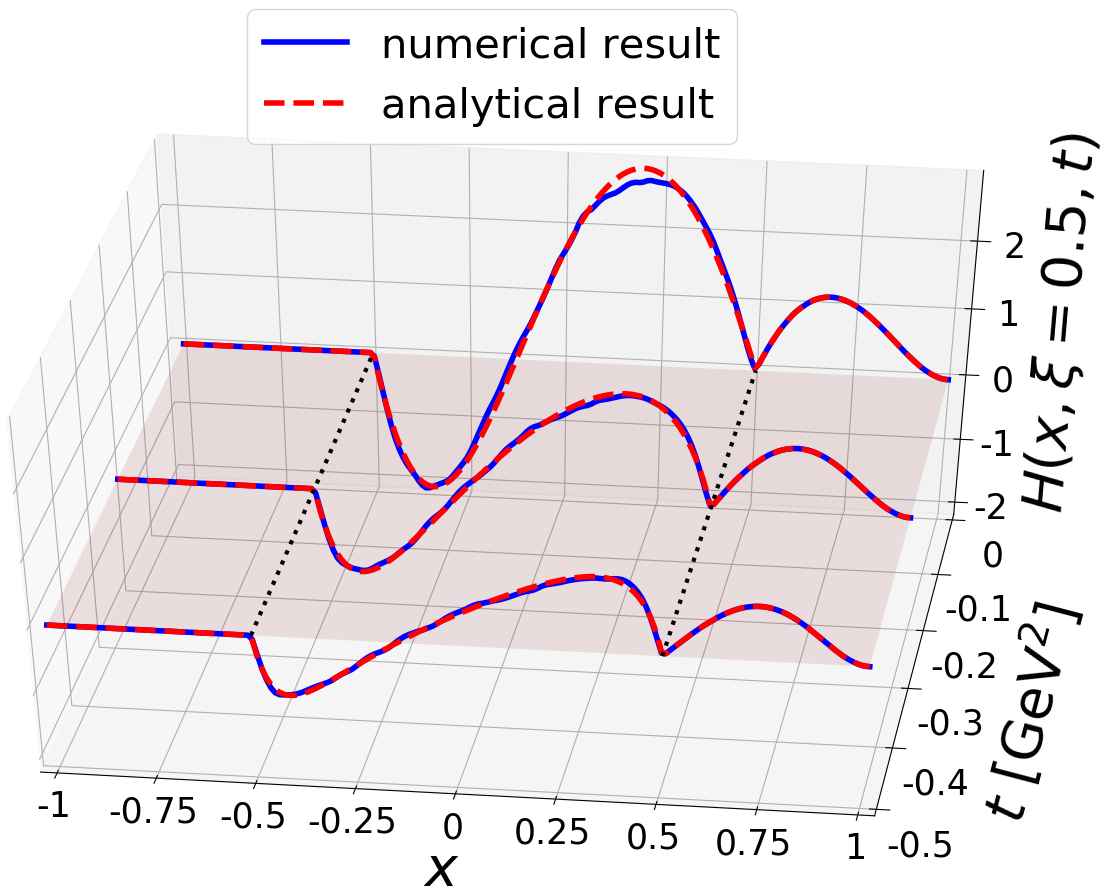}
  \quad
  \includegraphics[width = 0.4\textwidth]{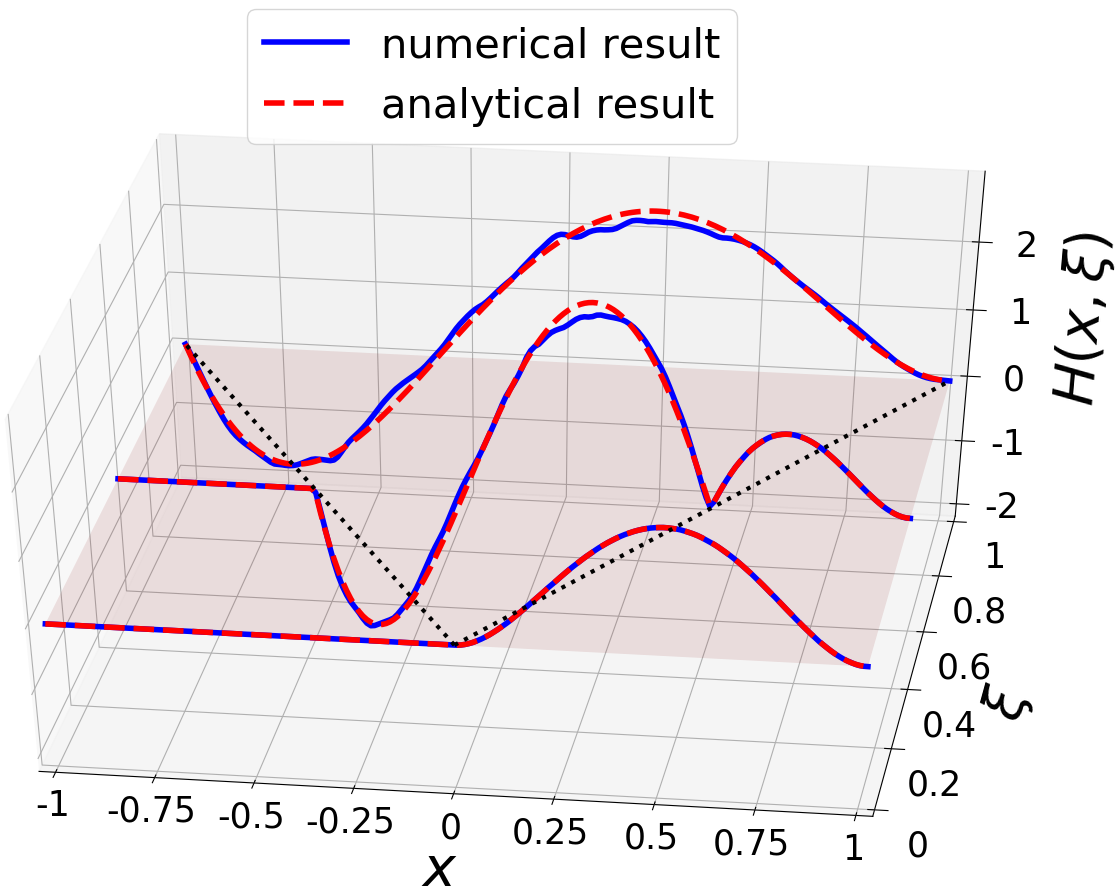}
  \caption{Example of reconstruction algorithm tested on the pion GPD model of \cite{Mezrag:2014jka,Mezrag:2014tva}. Red curve : algebraic answer to the inversion problem; Blue curve: numerical results.}
  \label{fig:Reconstruction}
\end{figure}

Our strategy has been tested on pion GPDs. One of the technical difficulty to overcome is the fact that the inverse Radon transform is ill-posed in the sens of Hadamard. That is, numerical noise must be handled with care. We developed an algorithm based on finite-element techniques which has been benchmarked using models whose inversion was known algebraically. An example is given in figure \ref{fig:Reconstruction} using a Nakanishi-based model \cite{Mezrag:2014jka,Mezrag:2014tva} for the pion GPD. Possible $D$-term like singularities can be partially fixed in this case using the so-called soft pion theorem \cite{Chouika:2017rzs}.

To conclude, we have presented a new method to model GPDs in such a way that all theoretical constraints are a priori guaranteed. The tests performed for the pion demonstrate the feasibility of our approach. The generalisation to the nucleon case is considered, especially since models of nucleon LFWFs are currently being developed (see e.g. \cite{Mezrag:2017znp}).

\subsection*{Acknowledgement}

This work is partly supported by the Commissariat à l’Énergie Atomique et aux Energies Alternatives, the GDR QCD “Chromodynamique Quantique”, the ANR-12-MONU-0008-01 “PARTONS” and the Spanish ministry Research Project FPA2014-53631-C2-2-P.

\bibliography{../../../../paper/Bibliography}
\bibliographystyle{JHEP}

\end{document}